# Disorder-driven carrier transport in atomic layer deposited ZnO thin films

**D. Saha[*], Amit. K. Das, R. S. Ajimsha, P. Misra, L. M. Kukreja**

Laser Materials Processing Division, Raja Ramanna Centre for Advanced Technology, Indore 452 013, India.

*Author for correspondence: babaisps@rrcat.gov.in

**Abstract:**

This paper addresses the effect of disorder on the carrier transport mechanism of atomic layer deposited ZnO thin films as has been investigated by temperature dependent electrical resistivity measurements in the temperature range of 4.2K to 300K. Films were grown on (00.1) sapphire substrate at different substrate temperatures varying from 150 to 350$^0$C. All the films were found to be degenerate in nature with carrier concentration exceeding the Mott's limit implying that films are on the metallic side of the metal-insulator transition (MIT). The defects and structural disorder in the films were found to be strongly dependent on their growth temperature. ZnO thin film with best optical and crystalline quality has been grown at 200$^0$C due to the self-limiting reaction chemistry between the precursor molecules on the growing film surface. However, at deposition temperatures lower (150$^0$C) and higher (250, 300 and 350$^0$C) than 200$^0$C ZnO thin film growth deviates from the ideal layer-by-layer growth mechanism of atomic layer deposition technique, leading to the growth of ZnO films of varying degree of disorder. The films grown at 150, 300 and 350$^0$C were found to be semiconductor-like in the whole measurement temperature range of resistivity ρ (T) due to the enhanced disorder in these films. However, a metal to semiconductor transition (MST) at low temperature has been observed in the films grown at 200 and 250$^0$C. It is also observed that the film grown at $250^0$C with higher residual resistivity, the transition temperature shifted towards the higher value due to the increased disorder in this film as compared to that grown at $200^0$C. The upturn in resistivity below the transition temperature has been well explained by considering quantum corrections to the Boltzmann's conductivity which includes the effect of weak localization and coulomb electron-electron interactions related to the existence of disorder in these films.

## I. Introduction

During the past few years, ZnO thin films and nanostructures have been extensively investigated due to their potential applications in emerging short-wavelength optoelectronic, transparent electronic and spintronic devices[1, 2]. So far, a variety of deposition techniques including pulsed laser deposition (PLD)[3], molecular beam epitaxy (MBE)[4] and chemical vapor deposition (CVD)[5] have been explored to grow ZnO thin films. Recently, atomic layer deposition (ALD)[6, 7], a particularly attractive technique in microelectronics manufacturing has also been employed to grow high-quality device grade ZnO thin films[8-11, 20-22]. ALD technique relies on two complementary and self-limiting surface reaction[6, 7] in which precursor molecules are exposed to the growing film surface in alternate pulses, separated by an inert gas purge. During the purging step the unreacted precursors and the reaction by-products are flushed away from the reactor. Unlike the conventional CVD technique in ALD there is no gas-phase reaction of the precursor molecules which facilitates the use of highly reactive and volatile precursors in ALD process. Due to the high reactivity of the precursors, ALD is used to grow thin films and nano structures at very low temperatures[6, 7]. ALD offers many advantages over other conventional thin film deposition techniques including its easy and precise thickness control down to molecular level, large area uniformity and excellent conformality on the underlying substrates, dense and pin hole free films and good reproducibility[6, 7]. Atomic layer deposited ZnO thin films have been found to be very promising as transparent conducting oxides (TCO)[8, 9] and as active channel layers in transparent thin film transistors (TFT)[10]. For efficient applications of ALD-ZnO films in advanced devices and further improvement and optimization of the TCO characteristics, it is very important to understand their charge transport mechanisms. It is reported in the existing literature that various native defects and thereby the carrier concentration in as grown ALD-ZnO thin films are strongly dependent on their growth temperature[9, 11]. In this context, we have grown ZnO thin films by ALD in a wide range of substrate temperatures varying from 150 to 350$^0$C keeping all the other process parameters fixed and studied their temperature dependent electrical resistivity from room temperature down to liquid-helium temperature to investigate the underlying carrier transport mechanisms. All the films were found to be n-type conducting and degenerate in nature with carrier concentration exceeding the Mott's limit[12]. The semiconducting behaviour and the presence of minima in temperature dependent resistivity curves has been well explained in the framework of quantum corrections to the Boltzmann conductivity[13-15] by taking into account the effect of weak localization and enhanced electron-electron interactions due to the presence of disorder in the ALD-ZnO thin films. The results of these studies are presented and discussed here.

**II. Experimental procedure**

ZnO thin films were grown on epi-polished (00.1) c-plane sapphire substrates by ALD (Beneq TFS-200) using diethylzinc (DEZ) and deionised water (H$_2$O) vapour as metal precursor and oxidant respectively. Substrate temperature was varied between 150$^0$C to

350$^0$C. Typical ZnO growth cycle by ALD consisted of 25 ms of DEZ exposure, 1s of N$_2$ purge, 25 ms of H$_2$O exposure, and 1s of N$_2$ purge in sequence. ZnO thin films of ~ 200 nm thickness were grown by adjusting the number of ALD cycles. Photo luminescence (PL) measurements were carried out using a 20 mW He-Cd laser operating at 325 nm as an excitation source and PL signals were collected and detected by a spectrometer (Triax 550, Jobin Yvon, France) attached with a UV sensitive CCD detector (Andor, UK). For electrical measurements, ohmic contacts were made on the films using Indium metal. The low temperature resistivity and Hall measurements were carried out in the temperature range from 4.2 to 300 K using the standard four-probe method in van der Pauw geometry with a magnetic field of 0.5 T.

**III. Results and discussions**

Fig. 1 shows the variation of room temperature resistivity (ρ), carrier concentration (n) and mobility (µ) of the ALD-ZnO thin films as a function of growth temperature. It can be seen from Fig. 1(a) that resistivity of the films decreased from 2.4x10$^{-2}$ Ω-cm to 3.8x10$^{-3}$ Ω-cm with increasing growth temperature from 150 to 200$^0$C and then increased with further increment in growth temperature. Fig. 1(b) shows that electron concentration of the films first increased from 1.6x10$^{19}$ to 7.8x10$^{19}$ cm$^{-3}$ as the growth temperature was increased from 150 to 200$^0$C and then decreased with further increase in growth temperature. The mobilities of the films (Fig. 1 (c)) were found to be almost independent of the growth temperature indicating that the change in resistivity was mainly due to change in carrier concentration in the films. This high level of electron concentration is already reported in literature for undoped ALD-ZnO films[8, 9, 11] and has been attributed to various intrinsic and extrinsic point defects such as zinc interstitials, oxygen vacancies, and/or hydrogen[16, 17]. These point defects introduce shallow donor levels below the conduction band minimum and are the sources of high electron doping in the ALD-ZnO films. Comparatively lower carrier concentration and higher value of room temperature resistivity of the films grown at 150, 300 and 350$^0$C results from the intensified disorder in these films as observed from the X-ray diffraction and room temperature photoluminescence (PL) measurements.

Figure 2 represents the room temperature PL spectra of all the ALD-ZnO thin films grown at different substrate temperatures. As can be seen the film grown at 200$^0$C showed intense excitonic near band edge emission (NBE) at ~ 380 nm corresponding to the band gap of ZnO (~3.3eV) and negligible deep level emission (DLE) in the visible spectral region[11]. The films grown at both lower (150$^0$C) and higher (250, 300 and 350$^0$C) substrate temperatures showed weak NBE and broad DLE[11]. The DLE intensity was found to be particularly strong for the films grown at 300 and 350$^0$C as compared to the films grown at 150 and 250$^0$C. It is reported that the broad DLE in ZnO is a superposition of different defect-related PL bands [18, 19]. The suppressed DLE and enhanced NBE in ZnO thin film grown at 200$^0$C indicated low defect density and good crystalline quality of the film. The inset of Fig. 2 showed the ratio of the peak intensities of NBE to that of DLE i.e., $I_{NBE}/I_{DLE}$ as function of the growth temperature. The ratio reaches a maximum at a substrate temperature of 200$^0$C implying that best optical quality of ALD-ZnO thin film was obtained at this temperature due the self-limiting reaction chemistry between the precursor molecules. However, the ratio gets reduced on either side of 200$^0$C indicating deterioration of their optical quality and enhanced defect formation in the films due to insufficient reactivity of the precursor molecules and their thermal decomposition or surface functional groups desorption[11, 20].

Fig. 3 shows the θ-2θ scans of all the ALD-ZnO thin films grown at different substrate temperatures. The XRD results revealed that ALD-ZnO films were polycrystalline in nature and their growth mode also changed with deposition temperature. As can be seen from Fig. 3, the films deposited at 200 and 250$^0$C of substrate temperatures showed preferred (00.2) oriented growth. However, the FWHM of the film grown at 250$^0$C was found to be higher as compared to the film grown at 200$^0$C indicating slight deterioration of the crystalline quality of the film grown at 250$^0$C. For the films grown at 150, 300 and 350$^0$C there was no preferred orientation of the crystallites. In these films diffused and weak XRD peaks were observed corresponding to different crystallographic planes of the hexagonal wurtzite ZnO. The films grown at 150 and 300$^0$C dominated by both (10.1) and (00.2) oriented crystallites. However, a negligible fraction of crystallites were found to be oriented along the (10.0) direction for the film grown at 350$^0$C. It is well studied that preferential growth of ALD-ZnO thin films is strongly dependent on their growth temperatures and other process parameters like precursor's pulsing and purging times etc[20-22]. Pung *et al.* reported that with changing deposition temperature the reaction chemistry between DEZ and H$_2$O might be changed due to the premature dissociation of DEZ molecules which resulted in changing of growth mode with varying crystalline quality[21]. In the present study ALD-ZnO thin films of different crystallographic orientation with varying crystalline quality has been grown by changing deposition temperature alone.

It is worthy to mention here that thermal decomposition of DEZ starts above 120$^0$C, which is typically not apparent during the growth of ALD-ZnO films on flat substrates due to its short pulsing time. The temperature above which precursors decomposition significantly contributes to the growth mechanism is strongly depends on the reactor design and process parameters[23]. In our case we observed a significant decrease in growth rate and deteriorated structural and optical quality of the ALD-ZnO films at 250$^0$C implying that DEZ decomposition starts affecting the quality of ALD-ZnO films at 250$^0$C which is more prominent in the films grown at 300 and 350$^0$C. However, the film grown at 150$^0$C showed high growth rate but poor optical and crystalline quality. The higher growth rate might be because of precursor condensation due to their insufficient reactivity to the surface functional groups. So we have a very narrow temperature window around 200°C where ZnO thin films with good structural and optical quality can be grown through the self-limiting reaction chemistry between the precursor molecules[11].

Results of temperature dependent resistivity in the temperature range from 300 K to 4.2K were shown in Fig. 4 for all the films deposited at different substrate temperatures in the range of 150-350°C. The curves 4 (a), 4 (d) and 4 (e), corresponding to the ALD-ZnO films grown at 150, 300 and 350°C respectively showed a semiconductor-like behaviour in the whole temperature range, down to liquid helium temperature. Whereas curves 4 (b) and 4 (c), corresponding to the films grown at substrate temperatures 200 and 250°C respectively exhibited a metallic conductivity above a certain transition temperature ($T_m$) and a semiconductor-like behaviour below it, i.e., a metal to semiconductor transition (MST) was observed in these films[13, 14]. The observed semiconductor-like parts in the resistivity curves of Fig. 4 (a), 4 (d), 4 (e) in the whole temperature range and Fig. 4 (b) and 4 (c) for $T < T_m$ could not be fitted by the classical Arrhenius equation

$$\rho(T) = \rho(0)\exp\left(E_a/KT\right) \quad (1)$$

$E_a$ is the activation energy to promote an electron into the conduction band. This suggests that carrier transport in these films is not due to simple thermal activation process as commonly observed in semiconductors. The above observation is not unexpected because the carrier concentrations in these films are quite high and they are degenerate in nature. Fig. 5 shows the temperature dependent hall measurement data for all the films. As can be seen from Fig. 5, for all the films the carrier concentration was nearly independent of temperature over the whole measurement temperature range, a characteristic of a degenerate electronic system[13]. For degenerate semiconductors a transition from an insulating to a metallic state (i.e., localized to itinerant electrons) occurs due to the formation of a degenerate band as suggested by Mott[24]. As can be seen from Fig. 1, the measured carrier concentrations for all the ALD-ZnO films exceed the Mott's critical carrier concentration for ZnO ($n_c \approx 5 \times 10^{18}$)[12]. In these heavily doped ZnO thin films Fermi level moves into the conduction band. However, despite their highly degenerate nature, semiconductor-like behaviour i.e., (d$\rho$/dT) < 0 was observed due to disorder induced carrier localization in these films[25, 26]. It is important to note that here we did not dope the films intentionally to create disorder in them as reported by Bhosle *et al*. in case of Ga doped ZnO films[25]. It is reported that undoped ZnO films can show MST due to the presence of various native defects and disorder in them[14]. Tiwari *et al*. showed a transition from the band-gap insulating state to the Anderson localized insulating state in oxygen deficit ZnO thin films due to the random distribution of oxygen vacancies which are responsible to create disorder and therefore localization of electronic states in the films[26]. The presence of native defects and structural disorder in our ALD-ZnO films can be observed from their room temperature PL spectra and XRD pattern as depicted in Fig. 2 and 3 respectively. The native donor type point defects increase the carrier concentration in the films and thereby reduce the electron correlation to Fermi energy ratio. However, these native point defects statistically occupy sites of the ZnO lattice and thus introduce disorder in the films leading to the localization of the delocalized electronic states at the Fermi energy. So both a Mott type (correlation) and an Anderson type (disorder) effects are prevalent as commonly observed in case of metal-insulator transition (MIT) of doped semiconductors. In this study we divide our samples into two categories. The first category consisted of the films grown at 200 and 250$^0$C which were within and slightly above the self-limiting growth window for ZnO deposition, showed a MST at low temperature. The second category consisted of the films grown at 150, 300 and 350$^0$C showed a semiconductor-like behaviour up to room temperature.

The observed semiconductor-like behaviour and the presence of minima in resistivity curves of Fig. 4(b) and 4(c) belonging to the first category films could be well explained in the framework of quantum corrections to Boltzmann conductivity[13-15, 25]. Such an approach can be implemented when the Fermi wavelength $\lambda_F$ ($2\pi/(3\pi^2 n_e)^{1/3}$) becomes comparable to the electronic mean free path $\Lambda$ (h/$\rho n_e e^2 \lambda_F$), where $n_e$ is the carrier density, $\rho$ is the resistivity and e is the electronic charge[15]. For ALD-ZnO thin films grown at temperatures 200 and 250$^0$C we obtained $\Lambda < \lambda_F$; hence, quantum corrections can be taken into account to interpret the transport mechanism in these films. The quantum corrections include the effect of weak localization and electron-electron interactions. The transport properties of these ALD-ZnO films were analyzed by fitting the experimental resistivity curves of the films to the equation[13-15]

$$\rho(T) = \frac{1}{\left(\sigma_0 + mT^{1/2} + aT^{p/2}\right)} + bT^2 \qquad (2)$$

Where $\sigma_0$ (1/$\rho_0$) is the residual conductivity, $aT^{p/2}$ corresponds to weak localization due to the self interference of coherent electron wave functions as the electrons are backscattered on impurities[15], in which p depends on the nature of interactions (p = 2 or 3 for electron-electron or electron-phonon interactions). The term $mT^{1/2}$ (Altshuler-Aronov correction) comes due to the long-range electron-electron interactions[15]. In addition to the quantum corrected resistivity terms, a term $bT^2$ is included in order to account for the high temperature scattering contributions. The resistivity curves in Fig. 4(b) and 4(c) show a decent fit to the above equation. The corresponding values of the fitting parameters are given in Table I. The best fit

(solid lines) for all the samples were obtained for p = 3 which means that electron-phonon interaction plays a dominating role in the carrier transport mechanism.

The above theory of quantum correction to the conductivity is valid provided that the quantum corrections are much smaller than the Boltzmann conductivity[14, 15], i.e.,

$$\delta\sigma = aT^{p/2} + mT^{1/2} << \sigma_0 \qquad (3)$$

Below the transition temperature ($T_m$) the values of $\delta\sigma/\sigma_0$ as calculated from the fitting parameters listed in TABLE-I for the ALD-ZnO films grown at 200 and $250^0$C were found to be 0.06 and 0.16 , i.e., quantum corrections are very small as compared to the Boltzmann conductivity. In general, quantum corrections are taken into account at very low temperatures to explain the anomalous temperature dependence of the electrical resistivity in degenerate semiconductors. However, in our ALD-ZnO thin films grown at substrate temperatures 200 and $250^0$C this effect remains dominant up to 130 and 190 K respectively. Such high values of transition temperature have been reported earlier in doped[25] and undoped ZnO[14] and in InGaN alloys[13]. It is important to observe that the film grown at $250^0$C has higher residual resistivity ($\rho_0$) and higher MST transition temperature ($T_m$) as compared to the film grown at $200^0$C. The increased value of $T_m$ is due to the enhanced disorder in the film grown at $250^0$C[25]. Varying degree of disorder in these two films can be explained by considering the reaction chemistry between the precursors at respective growth temperatures[11, 20]. At $200^0$C DEZ and H2O react with the surface functional groups in a self-limiting manner to grow layers of zinc and oxygen alternately. The layer-by-layer growth of thin films in a binary reaction sequence and self-limiting reaction between precursor molecules on the growing film surface which are the unique characteristics of ALD technique[6, 7] and essential for low defect formation[20] in the growing film has been achieved only for the film grown at $200^0$C. So defects and structural disorder in film grown at $200^0$C get reduced. However, at $250^0$C of substrate temperature ZnO film growth deviates from the self-limiting reaction chemistry due to thermal decomposition of DEZ, desorption of metal Zn from the growing film surface and surface dehydroxylation which leads to the enhanced defect formation in the films[11, 21].

In the second category, consisting of the films grown at 150, 300 and 350$^0$C of substrate temperatures have been found to exhibit semiconductor-like behaviour in the whole measurement temperature range. The increased value of their room temperature resistivity as compared to the films grown at 200 and 250$^0$C is due to the intensified disorder in these films. The enhanced defects in ALD-ZnO films grown at very low (150$^0$C) and very high (300 and 350$^0$C) deposition temperatures are highly possible. As discussed earlier at the lowest deposition temperature of 150$^0$C there might be incomplete reaction between the precursors due to insufficient thermal energy and at very high deposition temperatures, the films grown at 300 and 350$^0$C thermal decomposition of DEZ found to be more pronounced to enhance defect formation in the films[11, 20]. However, the degree of disorder in these films were not strong enough to turn them into Anderson localized insulators as observed by Tiwari et al[25]. Systems having carrier concentration less than $n_c$ will be on the insulating side of the metal-insulator transition and their resistivity curves are expected to follow Mott variable-range hopping[27] of the form

$$\rho(T) = \rho_0 \exp\left(T_0/T\right)^{1/4} \qquad (4)$$

The above equation is not fitting our experimental resistivity curves of the films grown at 150, 300 and 350$^0$C implying that carriers are not strongly localized in these films. The above observation is not unexpected because these highly disordered films are degenerate in nature with carrier concentration $n > n_c$. As it can be observed from Fig. 4(a), 4(d) and 4(e) that all these films were found to have a finite value of the residual conductivity ($\sigma_o$) and a very weak temperature dependence of resistivity with resistivity ratio ρ(r) = ρ (4.2K)/ ρ (300K) values 1.6-2.5 implying that the films are on the metallic side of the metal-insulator transition[27]. But they are showing semiconductor-like behaviour up to room temperature due to the enhanced disorder in these films which induces localization of the charge carriers.

**IV. Conclusion**

Temperature dependent carrier transport mechanisms of ALD-ZnO thin films of varying degree of disorder have been investigated on the metallic side of the metal-insulator transition. The films grown at150, 300 and 350$^0$C of substrate temperatures showed a semiconductor-like behaviour up to room temperature due to intensified disorder and poor crystalline quality of these films. However, the films grown at 200 and 250$^0$C exhibited a MST at low temperature due to reduced defects and better crystalline quality of these films. The film grown at 250$^0$C with higher residual resistivity shows the resistivity minimum at higher temperature, implying the significance of disorder in the low temperature carrier transport mechanism.

**Figure Captions**

**Figure 1.** Variation of (a) carrier concentration, (b) resistivity and (c) mobility of ALD-ZnO thin films with growth temperature.

**Figure 2.** Room temperature photoluminescence spectra of ALD-ZnO thin films grown on (00.1) sapphire substrate at growth temperatures of (a) 150, (b) 200, (c) 250, (d) 300 and (e) 350$^0$C. Inset shows the ratio of intensities of NBE to DLE peaks ($I_{NBE}/I_{DLE}$) as a function of growth temperature.

**Figure 3.** XRD pattern of ALD grown ZnO thin films on (00.1) sapphire substrate at growth temperatures of (a) 150, (b) 200, (c) 250, (d) 300 and (e) 350$^0$C.

**Figure 4.** Variation of resistivity with measurement temperature for ALD-ZnO thin films grown at substrate temperatures of (a) 150, (b) 200, (c) 250, (d) 300 and (e) 350$^0$C.

**Figure 5.** Variation of carrier concentration with measurement temperature for ALD-ZnO thin films grown at substrate temperatures of 150, 200, 250, 300 and 350$^0$C.

**TABLE I.** Values of the fitting parameters of Eq. (2) for the ALD-ZnO thin films grown at substrate temperatures 200 and 250$^0$C on (00.1) sapphire substrate.

| Growth temperature ($^0$C) | $\sigma_0$ ($\Omega^{-1}$cm$^{-1}$) | m ($\Omega^{-1}$cm$^{-1}$K$^{-1/2}$) | a ($\Omega^{-1}$cm$^{-1}$K$^{-1}$) | b ($\Omega$ cm K$^{-2}$) |
|---|---|---|---|---|
| 200$^0$C | (270.3±0.1) | (4.8±0.8)x10$^{-2}$ | (1.08±0.01)x10$^{-2}$ | (8.28±0.05)x10$^{-9}$ |
| 250$^0$C | (146.2±0.1) | (6.8±0.5)x10$^{-2}$ | (8.59±0.07)x10$^{-3}$ | (1.55±0.01)x10$^{-8}$ |

**Figure 1:**

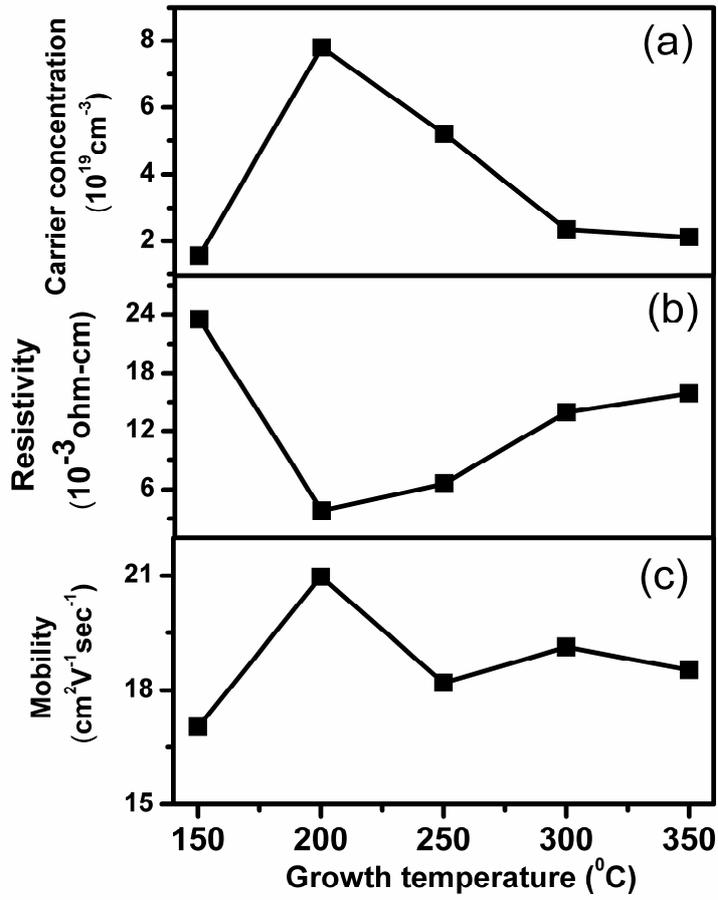

**Figure 2:**

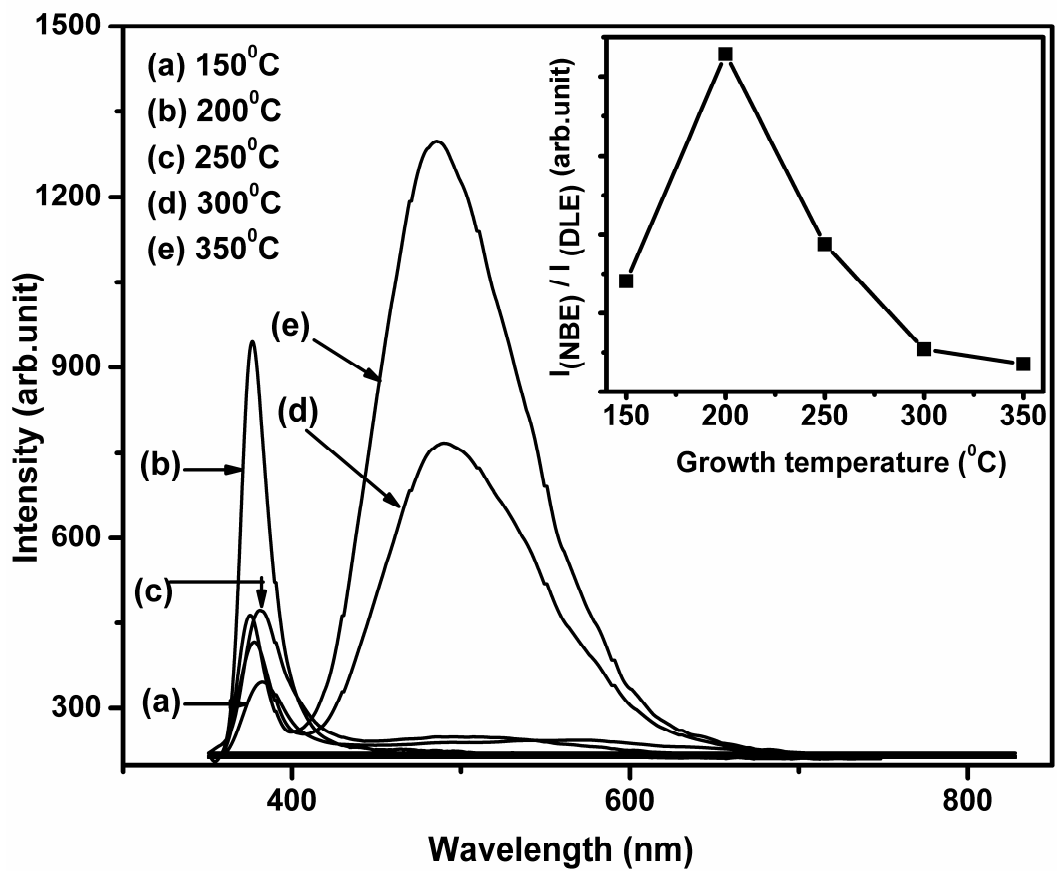

**Figure 3:**

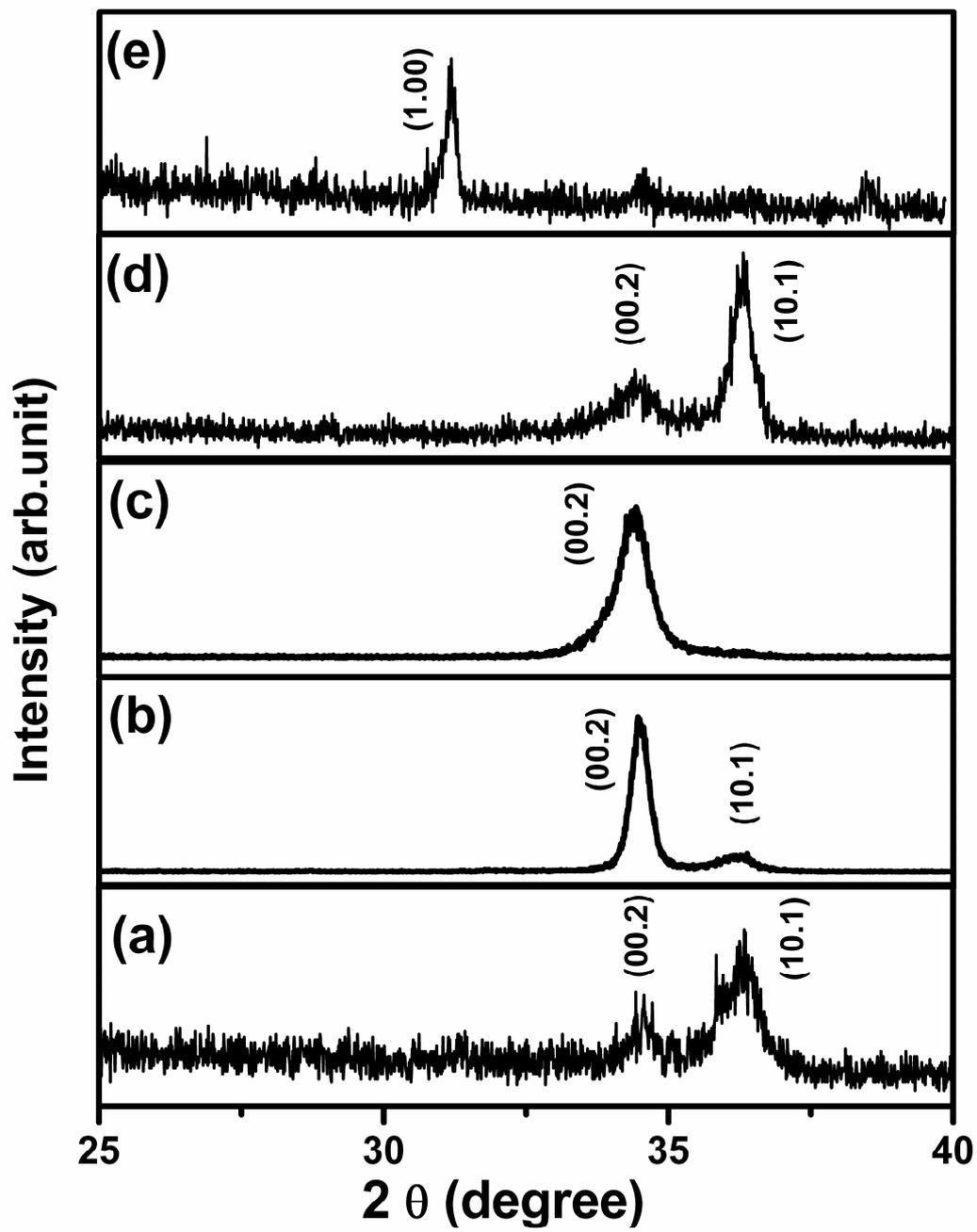

**Figure 4:**

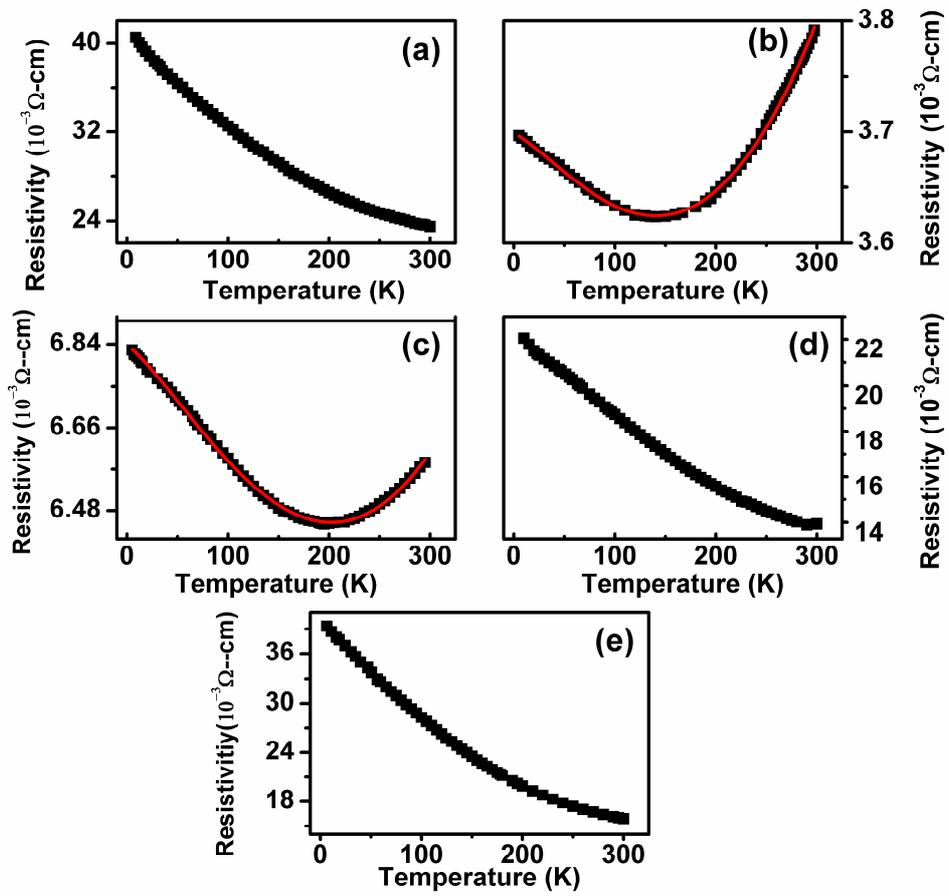

**Figure 5:**

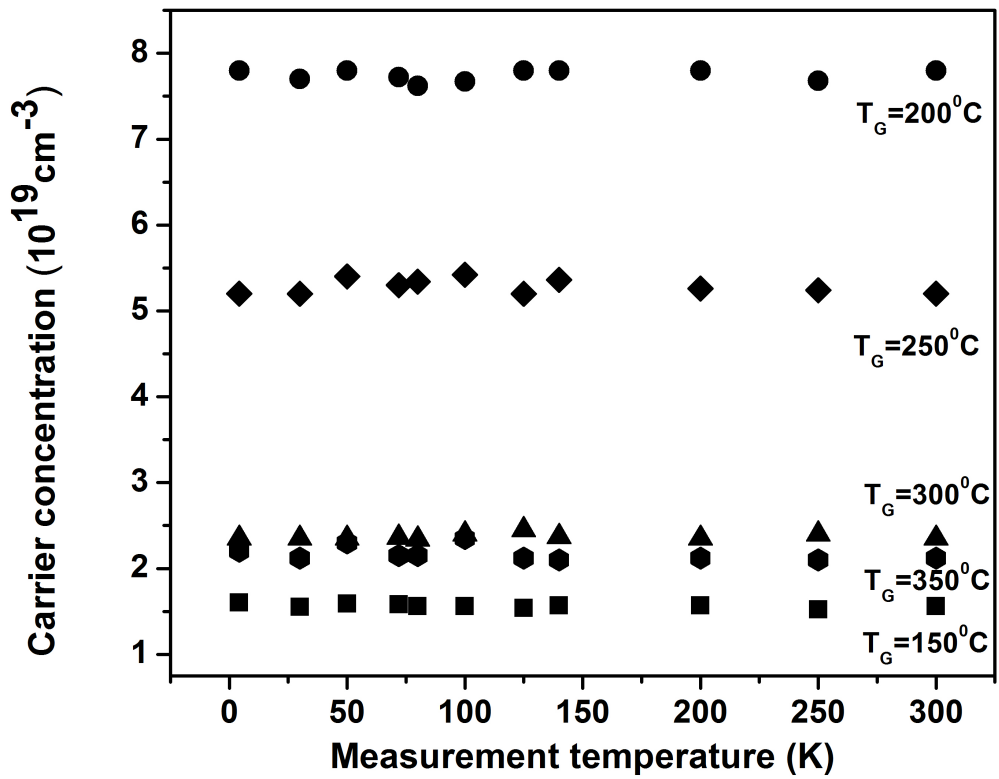